\documentclass[10pt, conference, letterpaper]{IEEEtran}  
\usepackage{url,comment,cite}
\usepackage{amsmath,amssymb,amsfonts,bbm}
\usepackage{graphicx}
\usepackage{epsfig,epsf,psfrag,textcomp,tabularx}
\usepackage[usenames,dvipsnames]{xcolor}
\usepackage{subfigure}
\usepackage{float}
\usepackage{stfloats,mathtools}
\usepackage[algoruled,medskip,dontprintsemicolon,linesnumbered,Algorithm]{algorithm}
\usepackage[noend]{algpseudocode}

\newtheorem{theorem}{Theorem}

\newcommand{\Cc}{\mathcal{C}}

\newcommand{\Lc}{\mathcal{L}}
\newcommand{\Pc}{\mathcal{P}}

\newcommand{\Wc}{\mathcal{W}}
\newcommand{\Tc}{\mathcal{T}}

\newcommand{\Sc}{\mathcal{S}}
\newcommand{\Zc}{\mathcal{Z}}

\numberwithin{equation}{section}

\begin{document}

\title{Downlink  Transmit Power Setting \\in LTE HetNets with Carrier Aggregation}

\author{\IEEEauthorblockN{Zana Limani Fazliu, Carla Fabiana Chiasserini}
\IEEEauthorblockA{Politecnico di Torino, Italy} 
\and
\IEEEauthorblockN{Gian Michele Dell'Aera}
\IEEEauthorblockA{Telecom Italia, Italy}}

\maketitle
\thispagestyle{plain}
\pagestyle{plain}

\begin{abstract}
Carrier aggregation, which allows users to aggregate several component carriers to obtain up to 100 MHz of bandwidth, is one of the central features envisioned for next generation cellular networks. While this feature will enable support for higher data rates and improve quality of service, it may also be employed as an effective interference mitigation technique, especially in multi-tier heterogeneous networks. Having in mind that the aggregated component carriers may belong to different frequency bands and, hence, have varying propagation profiles, we argue that it is not necessary, indeed even harmful, to transmit at maximum power at all carriers, at all times. Rather, by using game theory, we design a distributed algorithm that lets eNodeBs and micro base stations dynamically adjust the downlink transmit power for the different component carriers. We compare our scheme to different power strategies combined with popular interference mitigation techniques, in a typical large-scale scenario, and show that our solution significantly outperforms the other strategies in terms of global network utility, power consumption and user throughput.
\end{abstract}

\section{\label{sec:intro}Introduction} 
The exponential increase in mobile data traffic in recent years has become a serious challenge for today's cellular communication networks. 
To tackle this challenge, one of the strategies foreseen in the LTE-Advanced (LTE-A) specifications, among others,  is the deployment of Heterogeneous Networks (HetNets). HetNets are seen as a potential cost-efficient approach to effectively meet the challenge, by introducing smaller cells, i.e., micro, pico and femtocells, nested within the traditional macrocell. This approach promises to improve both the capacity and the coverage of current cellular networks. However, it also introduces several technical challenges, the most prominent being the interference between different architectural layers sharing the same spectrum resources. 

Carrier aggregation is another expected feature of future networks, which aims at guaranteeing higher data rates for end users so as to meet the IMT-Advanced requirements. It enables the concurrent use of several LTE component carriers with,  potentially, different bandwidth 
and belonging to different frequency bands. Downlink transmissions over each carrier will occur at maximum output power and each carrier will have an independent power budget \cite{3gpp-trca}. Thus, different component carriers may have very different coverage areas and impact in terms of interference, due to both their different transmit power level and  propagation characteristics. 

Currently, three main approaches have been proposed  to mitigate downlink interference in HetNets: per-tier assignment of carriers, Enhanced Inter Cell Interference Coordination (eICIC), which has  been  adopted in LTE-A systems, and downlink power control. Per-tier assignment of carriers simply implies that in HetNets with carrier aggregation support, each tier should be assigned a different carrier component so as to nullify inter-tier interference \cite{lp-abs}. eICIC includes techniques such as Cell Range Expansion  to incentivise users to associate with micro base stations (BSs), and Almost Blank Subframes (ABS), i.e.,  subframes during which macro BSs mute their transmissions to alleviate the interference caused to microcells.   Algorithms to optimise  biasing coefficients and ABS patterns in LTE HetNets have been studied in, e.g., \cite{eicic-alg}, however they do not address carrier aggregation.  Also, modifications to the eICIC techniques that allow macro BSs to transmit at reduced power during ABS subframes have been proposed in \cite{lp-abs}. In this paper we do not consider a solution within the framework of eICIC or its modifications, rather we use them as comparison benchmarks for the solutions we propose. 
We adopt instead the third approach, which consists in properly setting the downlink transmit power of BSs so as to avoid interference between different tiers. Indeed, macro BSs transmitting blindly at high power ensures large coverage  and an acceptable level of service for all users under coverage, but it can also create a lot of harmful interference to  microcell users. Interesting schemes adopting a similar approach 
have been proposed in \cite{coalitions_overlap,hierarchical-competition,discrete-eeff}, which, however, did not consider carrier aggregation. 

In this paper we address the problem of downlink power setting in LTE HetNets with carrier aggregation support, when all BSs share the available radio resources. Carrier aggregation allows all carrier aggregation enabled users in the network to receive concurrently in two or more component carriers, while they are under their coverage areas.  The coverage area of each component carrier is determined by the carrier's propagation characteristics, as well its transmitting power, therefore it is possible that some users may be under the coverage of one carrier and not others. We propose to leverage this diversity  in the component carrier coverage areas to mitigate inter-tier interference in HetNets. In addition, by varying the carrier transmit power to alter their coverage, we enable a wide range of network configurations which reduce power consumption, provide high throughput and ensure a high level of coverage to network users. This type of configurations have also been envisioned by 3GPP \cite{3gpp-ca}, however, unlike the current specifications, we aim at reaching such solutions dynamically and in-response to real traffic demand.

As envisioned in LTE-A systems, we consider that each component carrier at each BS has an independent power budget, and that BSs can choose the transmit power on each carrier from a set of discrete possible values.   This implies that the problem we face is not to properly allocate the power among the different carriers to ensure most efficient use of a power budget, as done in existing work. Rather, we address the problem of adequately choosing a power level from a range of choices to ensure optimal network performance. 
It is easy to see that the complexity of the problem increases exponentially with the number of cells, carriers and the granularity of the power levels available to the BSs. In addition, 
if one of the objectives is to maximise the network throughput, the problem becomes  
non linear  since transmission data rates depend on the signal-to-interference-plus-noise ratio (SINR) experienced by the users. It follows that  
an optimal solution requiring a centralised approach would be both unfeasible and unrealistic, given the large number of cells in the network and the flat network architecture of LTE-A. 

We therefore study the above problem through the lens of game theory, which is 
 an excellent mathematical tool to obtain a multi-objective distributed solution in a scenario with entities (BSs) sharing the same pool of resources (available component carriers). We model each group of BSs in the coverage area of a macro BS as a team so that we can capture both (i)  cooperation between  macro BS and  micro BSs with overlapping coverage areas, and (ii) the competitive interests of different macro BSs. The framework we provide however allows for straightforward extension to teams which include several macro BSs. 
We prove that the game we model belongs to the class of {\em pseudo-potential} games, which are known to admit pure Nash Equilibria (NE) \cite{pa-potential}. This allows us to propose a distributed algorithm based on best-reply dynamics that  enables the network to dynamically reach an  NE representing the preferred solution in terms of throughput, user coverage and power consumption. As shown by simulation results, our scheme outperforms fixed transmit power strategies, 
even when advanced interference mitigation techniques such as eICIC are employed.

\section{\label{sec:rel-work}Related work} 
We focus our discussion on existing works on power control in cellular networks, since they are the most relevant to our study. Note that, while many papers have appeared in the literature on uplink power control, fewer exist on downlink power setting.  
Among these,  
\cite{coalitions_overlap}  uses  coalitional games to investigate  power and resource allocation in HetNets where cooperation between players is allowed.  Downlink power allocation in cellular networks 
is modeled  in  \cite{hierarchical-competition} as a Stackelberg game, with macro and femto BSs competing to maximise their individual capacities under power constraints. 

An energy efficient approach is instead proposed in \cite{hetnet-eff}. There, BSs do not select transmit power levels as we do in our work, rather they can only choose between on and off states. 
Maximising  energy efficiency is also the goal of \cite{yang-eeff}, which however is limited  to the study of resource allocation and downlink transmit power in a two-tier LTE single cell. 
A multi-cell network with inter-cell interference is considered in \cite{discrete-eeff}, where   energy efficiency is optimised by applying resource allocation and discrete transmit power levels. 
We remark that the above papers address HetNets but, unlike our work, they do not consider carrier aggregation support. Also, \cite{yang-eeff, coalitions_overlap, hierarchical-competition}  formulate a resource allocation problem that aims at distributing the transmit power among the available resources under overall power constraints. In our work, instead, 
we do not formulate the problem as a downlink power allocation problem,  rather as a  power setting problem at carrier level, assuming {\em each carrier has an independent power budget}. Additionally, while most of the previous work \cite{hetnet-eff, discrete-eeff, yang-eeff} focus on the HetNet interference problem only, using game theory concepts we  jointly address  interference mitigation, power consumption  and user coverage by taking advantage of the diversity and flexibility provided by the availability of multiple component carriers.  Finally, we  propose a 
solution that enables the BSs to dynamically change their power strategies based on user distribution, propagation conditions and traffic patterns. 

To our knowledge, the only existing work that investigates downlink power setting in LTE networks with carrier aggregation support is  \cite{joint-ra-ca}.  
There, the authors formulate an optimisation problem that aims at maximising  the system energy efficiency by optimising power allocation and user association. However, interference issues, which are one of the main challenges we address, are largely ignored in \cite{joint-ra-ca} as the authors  consider a non-heterogeneous single cell scenario.


\section{System model and assumptions\label{sec:system}}
We consider a two-tier LTE network composed of macro BSs controlling  macrocells, and 
micro BSs controlling microcells. 
For simplicity, the user equipments (UEs) in the network area are all assumed to be carrier aggregation (CA) enabled. 
Note, however, that the extension to  
a higher number of tiers as well as to the case where there is a mix of CA-enabled and non CA-enabled UEs is straightforward.

The network area is partitioned into a set of tiles, or zones, denoted by $\Zc$. From the perspective of downlink power setting, all UEs within a tile $z \in \Zc$ are assumed to experience the same propagation conditions from a specific BS. Also, the tile-BS association is determined by the mobile operator network planning. In particular, following~\cite{qualcomm}, we will assume for ease of presentation that tiles (i.e., the UEs therein) are associated with the closest BS, although the extension to other, dynamic association schemes as well as to the case where a tile is served by multiple BSs can be easily  obtained. 

All BSs share the same radio resources. In particular, a comprehensive set of component carriers (CC), indicated by $\Cc$, is available simultaneously at all BSs (BSs having at their disposal a subset of CCs is a sub-case of this scenario). 
Each CC is defined by a central frequency  
 and a certain bandwidth.  The central frequency affects the carrier's coverage area, as the propagation conditions deteriorate greatly with increasing frequency.
The level of transmit power irradiated by each BS on the available CCs can be updated periodically depending on the traffic and propagation conditions in the served  tiles, or it can be triggered by changes in such network parameters. The update time interval, however, is expected to be substantially longer than a resource block allocation period, e.g., order  of hundreds of subframes. The BSs can choose from a discrete set of available power levels, including 0 that corresponds to switching off the CC. The possible power values are expressed as fractions of the maximum transmit power, i.e., $\boldsymbol{P}=\{0.1, 0.2,...,1\}$, with the  maximum transmit power that typically depends on the type of BS. As noted before, each CC at each BS has an independent power budget. 

In order to determine the downlink power setting, BSs can  leverage the feedback  they receive from their users on the channel quality that UEs experience. Also, we assume that each macro BS is connected with the set of micro BSs underlaid over its coverage area,  via, e.g.,  optical fiber connections, which allows for swift communication 
between them.  As a result, we assume that it is possible for the macro BS and the corresponding micro BSs to cooperate and exchange information in order to reach common decisions. This is a reasonable assumption since it is expected that the architecture foreseen for future networks will allow BSs that are geographically close to share a common baseband~\cite{ericsson}. Furthermore, it is fair to assume that neighbouring macro BSs can   communicate with each other.




\section{Game theory approach\label{sec:game}}

\begin{figure}
\psfrag{M1}[c][]{\footnotesize{$l_1$}}
\psfrag{M2}[c][]{\footnotesize{$l_1$}}
\psfrag{M3}[c][]{\footnotesize{$l_1$}}
\psfrag{P1}[c][]{\footnotesize{$l_2$}}
\psfrag{P2}[c][]{\footnotesize{$l_3$}}
\psfrag{P3}[c][]{\footnotesize{$l_2$}}
\psfrag{P4}[c][]{\footnotesize{$l_3$}}
\psfrag{P5}[c][]{\footnotesize{$l_2$}}
\psfrag{P6}[c][]{\footnotesize{$l_3$}}
\psfrag{T1}[c][]{\footnotesize{{\bf Team 1}}}
\psfrag{T2}[c][]{\footnotesize{{\bf Team 2}}}
\psfrag{T3}[c][]{\footnotesize{{\bf Team 3}}}
\centering
	\includegraphics[width=0.5\columnwidth]{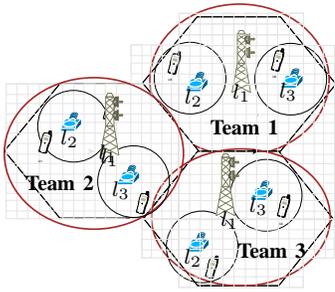}
\caption{\label{fig:net-model}Network model and teams. 
Team locations are denoted by $l_1, l_2, l_3$. Solid red lines represent team boundaries, while black solid lines represent coverage areas. Tiles are represented by grey squares.}
\vspace{-5mm}
\end{figure}
As mentioned, the complexity of  carrier power setting   
may be very high and impair an optimal, centralised solution in networks 
with many cells. 
We therefore  adopt a  game theoretic approach to the problem, which provides a low-complexity, distributed solution that is  applicable in realistic scenarios.


We formulate the problem of power setting in LTE HetNets with carrier aggregation as a competitive game between {\em teams} of BSs (see Fig.~\ref{fig:net-model}),  where each team wants to maximize its own payoff. Indeed, given the network architecture at hand,  a macro BS and the micro BSs within its coverage area, have the common objective to provide the UEs located within the geographical area of the  macrocell with a high data throughput. Thus, they  may choose to cooperate with each other in order to improve their individual payoffs as well as contribute to the ``public good'' of the team. Cooperation between such BSs is beneficial especially since the inter-tier interference is most significant within the cell. 
 At the same time, although increasing the transmit power of one BS may increase the SINR that its UEs experience,  
such increase  hurts the UEs being served by other BSs since all BSs  share the same frequency spectrum. It follows that teams will compete between each other for the same resources, each aiming at  maximising its own benefits. 

The game we model  and its analysis are detailed below.


\subsection{Game definition\label{subsec:game-definition}}
Let $\Tc=\{t_1,...,t_{T}\}$ be the set of teams in our network, where $T$ is the number of teams. Each team consists of a macro BS and  the micro BSs whose coverage areas geographically overlap with that of the macro BS.  Note that not only can team players  exchange information between each other, but we can also assume that the macro BS  plays the role of  team leader, i.e., it makes the decisions for all team members in a way that maximizes the overall team benefits. 

To generalise the formulation further, we will refer to the BSs forming a team $t$ as the {\em locations} of the team,  $\Lc_t=\{l_1,l_2,...,l_L\}$ where, for simplicity of notation, the number of locations within a team is assumed to be constant and equal to $L$. Such a generalisation is particularly useful since the interference caused within the team depends also on the relative position between the different players. We indicate the set of tiles under the coverage area of a particular location $l$ by $\Zc_l$, and their union, denoting the comprehensive set of tiles of the team, by $\Zc_t$.  
Also, let us denote by $E_l$ the number of UEs under the coverage of location $l$, and by $E_t=\sum_{l\in\Lc_t}E_l$  the total number of UEs served by the team. 

Each team, comprising  a set of locations (BSs located at different positions within the macrocell), has to decide which  transmit power level to use (out of the possible values in $\boldsymbol{P}$), at each one of those locations and for each of the available carriers $\Cc=\{c_1,c_2,...,c_C\}$. 
It follows that the strategy selected by a team $t$, $\boldsymbol{s^t}$, is an $L\times C$ matrix, where each $(l,c)$ entry indicates the power level set at location $l$ on carrier $c$. 

We now provide the definitions for the team utility and payoff, which are used in game theory to model the objectives of the players when choosing their strategy. Since network throughput is an important performance metric, it is natural that the utility of each team is defined as a function of the data rates it can serve to its UEs. The data rate a UE obtains is closely linked to the SINR it experiences, which depends on the transmit power chosen by the serving location (BS), the CC that is used and the transmit power levels chosen by neighbouring locations. Assuming that all UEs within the same tile experience the same amount of interference, for each team we can first define an interference matrix of size $|\Zc_t| \times C$, denoted by $\boldsymbol{I^t}$.  Each entry in the matrix indicates the interference experienced by UEs in tile $z$ on carrier $c$, which is caused by other teams:  
\begin{equation}
I^t_{z,c}(\boldsymbol{s^{-t}}) = \sum_{t'\in\Tc \wedge t'\neq t}\sum_{l'\in\Lc_{t'}}s^{t'}_{l',c}a_{l',z,c}
\label{eq:interference}
\end{equation}
where $\boldsymbol{s^{-t}}$ represents the strategies adopted by all teams other than $t$, $s^{t'}_{l',c}$ is the power level (the strategy) of team $t'$ for location $l'$ on carrier $c$ and $a_{l',z,c}$ is the factor of the attenuation ($0\leq a_{l',z,c}\leq1$) experienced by the signal transmitted from location $l'$ on $c$ when it reaches the UEs in tile $z$. The attenuation values are pre-calculated using the urban propagation models specified in \cite{itu}.

The SINR  at tile $z$, when served by location $l$ in team $t$, is:
\begin{equation}
\gamma_{z,c}^t=\frac{s^{t}_{l,c}a_{l,z,c}}{N+\sum_{l'\in\Lc_t \wedge l'\neq l}a_{l',z,c}s^t_{l',c}+I^t_{z,c}}
\label{eq:SINR}
\end{equation}
where $N$ represents the average noise power level. Note that, besides $N$ and $I^t_{z,c}$, we have an additional term in the denominator,  which stands for the intra-team interference and indicates the sum of all power received from the locations within the same team, other than  location $l$. 

Then the utility of each team can be defined as a function of the individual tiles' SINR values. 
In particular, the sigmoid-like function has been  often used for this purpose in uplink power control \cite{pa-sigmoid}. We note that this function is  suited to capture also the utility in  downlink power setting, as it has features that closely resemble the realistic relationship between the SINR and the data rate.  
We therefore adopt the sigmoid function proposed in \cite{pa-sigmoid}, as the utility function of each (tile, carrier) duplet  in the team, and write the team utility as:
\begin{equation}
u^t(\boldsymbol{s^t},\boldsymbol{s^{-t}}) = \sum_{l\in\Lc_t} \sum_{z\in\Zc_l}\sum_{c\in\Cc}\frac{E_z}{E_t \left(1+e^{-\alpha(\gamma^t_{z,c}-\beta)} \right)} \,. \label{eq:team-utility-sigmoid}
\end{equation}
The sigmoid function in Eq.~(\ref{eq:team-utility-sigmoid}) has two tuneable parameters, $\alpha$, which controls the steepness of the function, and $\beta$, which controls its centre.  They can be tweaked to 
best meet the scenario of interest. In particular, the higher the $\alpha$, the closer the function resembles a step function, i.e., the utility becomes more discontinuous with the increase of the SINR. The higher the $\beta$, the larger the  SINR for which a tile  obtains a positive utility (see Sec.~\ref{sec:peva} for the setting of these parameters). Also, 
the individual utility of each tile $z$ in team $t$ is weighted by the fraction of UEs
 covered by the team  in the tile ($E_z/E_t$) so as to give more weight to more populated tiles. This enables us to account for the user spatial distribution whenever this is not uniform over the network area.  

Next, we introduce a cost function to account for the interference and its detrimental effect, as well as for fairness in the service level to users.  We define a first cost component that aims at penalising  players who choose high power strategies, as:
$\xi \sum_{l\in\Lc_t}\sum_{c\in\Cc}\bar{a}_{l,c}s^t_{l,c}$ 
where $\bar{a}_{l,c}$ is the link quality on carrier $c$ averaged over all  tiles served by location $l$, and  $\xi$ is the unit price per received power. This cost component increases with the increase in the chosen level of transmit power, however  it also accounts for the propagation conditions of the users served by the location. 
In other words, locations that have to serve UEs experiencing poor channel quality will incur a lower  cost, which ensures some level of fairness. Additionally, as clear by intuition and as shown in our technical report~\cite{techpaper}, the parameter $\xi$ can be optimally set so as to be inversely proportional to the average interference that the team experiences from other teams. This way the cost component will be smaller for a team that experiences high interference thus rightfully pushing the team to  increase  its transmit power.

The second term of the cost function further provides fairness in the network 
by  penalising those strategies that leave  UEs without coverage. It is defined as
$\delta e_t$, 
where $\delta$ is a unit price paid for each unserved user and $e_t$ is the fraction of UEs within the team area that experience SINR levels below a certain threshold. We remark that since a macro BS can communicate with the  micro BSs in the macrocell, the team leader has knowledge of the UE density  under the coverage of its team players. Thus, it can easily estimate the fraction of users, $e_t$, depending on the strategy  chosen for each of its players ($\boldsymbol{s^t}$) as well as  on all other teams' strategies ($\boldsymbol{s^{-t}}$).  
The total  cost function is then given by: 
\begin{align}
\pi^t(\boldsymbol{s^t},\boldsymbol{s^{-t}})= \xi\sum_{l\in\Lc_t}\sum_{c\in\Cc}\bar{a}_{l,c}s^t_{l,c}+\delta e_t \label{eq:fullcost}
\end{align}
%
where $\xi$ and $\delta$  indicate the weight that is assigned to each part of the cost function. 
Finally, we define the payoff of each team $t$ as the utility minus the cost paid:
\begin{align}
w^t(\boldsymbol{s^t},\boldsymbol{s^{-t}})  = u^t(\boldsymbol{s^t},\boldsymbol{s^{-t}}) -\pi^t(\boldsymbol{s^t},\boldsymbol{s^{-t}})  \,.\label{eq:teampayoff}
\end{align}

In summary, we can formulate the problem as a competitive game $G=\{\Tc,\Sc,\Wc\}$, where $\Tc$ is the set of teams, $\Sc$ is the comprehensive set of  strategies available to the
teams, and $\Wc$ is the set of payoff functions. The objective of each team is to choose a strategy that maximises its payoff. Because its payoff depends also on the strategies of the other teams, a team must make decisions 
accounting for the strategies, it estimates or knows,  the other teams have selected. Thus, using game-theory terminology, we will refer to the strategy chosen by a team as best reply. Moreover, to reduce both power consumption and the interference towards other teams, a team will select its best reply among  strategies that maximise its payoff, as follows. \\
\noindent
{\em (i)}~Between strategies that are equivalent in terms of payoff, it will choose the one with the lowest total power, to reduce the overall power consumption. \\
\noindent
{\em (ii)}~When indifferent between  strategies with equal total power but assigned to different locations, it will select the strategy that  assigns higher power levels to micro BSs that are closer to the centre of the cell, to minimise interference. \\
\noindent
{\em (iii)}~When indifferent with respect to the two above criteria,
it will choose the strategy that assigns higher power levels to higher frequency carriers, again, to minimise interference. 

\subsection{Game analysis\label{subsec:game-analysis}}
To analyse the behaviour of the above-defined game, and discuss the existence of NEs, we rely on the definition of games of {\em strategic complements/substitutes with aggregation} as provided in~\cite{pa-potential,pa-strategic}. 

A game $\Gamma=\{\Pc,\Sc,\Wc\}$, where $\Pc$ is the set of players, and $\Sc$ and $\Wc$ are defined as above, is a game of {\bf{strategic substitutes}} with aggregation if for each player 
 $p\in \Pc$ there exists a best-reply function  $\theta_p:\boldsymbol{S^{-p}} \to \boldsymbol{S^p}$ such that:
\begin{align}
1)& \theta_p(I^p)\in \Theta(I^p)\label{eq:cond-1}\\ 
2)& \theta_p\text{ is continuous in } \boldsymbol{S^{-p}}\label{eq:cond-2} \\
3)& \theta_p(\hat{I}^p) \leq \theta_p(I^p), \forall \hat{I}^p>I^p \,.\label{eq:cond-3}
\end{align}
$\Theta(I^p)$ is the set of best replies for player $p$  and $\boldsymbol{S^{-p}}$ is the Cartesian product of the strategy sets of all participating players other than $p$.  $I^p$ is an additive function of all other players' strategies, also referred to as the {\em aggregator} \cite{pa-strategic}: 
\begin{equation}
I^p(\boldsymbol{s^{-p}}) =\sum_{p'\in\Pc, p' \neq p} b_{p'}s_{p'}\label{eq:aggregator}
\end{equation}
where $b_{p'}$ are scalar values.  
Condition 1) is fulfilled whenever the dependence of the payoff function on the other players' strategies can be completely encompassed by the aggregator. Condition 2), also known as the {\em continuity} condition, implies that for each possible value of $I^p$, the best reply function $\theta_p$ provides unique best replies.  Condition 3) implies that the best reply of the team decreases with the value of the aggregator. 

A game of {\bf{strategic complements}} with aggregation is identical, except for condition 3), which changes into:
\begin{align}
\theta_p(\hat{I}^p) \leq \theta_p(I^p), \forall \hat{I}^p<I^p \,,\label{eq:cond-4}
\end{align}
i.e., in the case of games of strategic complements, the best reply of the team increases with the value of the aggregator.

Next, we show the following important result. 
\begin{theorem}
{\em Our competitive team-based game $G$ 
is a game of {\bf strategic complements/substitutes with aggregation}. } 
\end{theorem}
\IEEEproof
For brevity, here we provide a sketch of the proof; the full proof can be found in~\cite{techpaper}. Let us define the aggregator in our scenario as the interference experienced by a team. It is easy to see that, in the case of a single-player team and a  single carrier, such aggregator satisfies condition (\ref{eq:aggregator}), and that   
$G$ meets the conditions set out in Eqs.~(\ref{eq:cond-1})-(\ref{eq:cond-2}) and in either  Eq.~(\ref{eq:cond-3}) or Eq.~(\ref{eq:cond-4}). The extension to a multi-carrier game with  multi-player teams, implies that the strategy chosen by the team is not a scalar value but a matrix. Likewise,  the interference experienced by each team (i.e., the  aggregator) is a matrix. 
Given that, and similarly to the scalar case, it can be verified that the team best reply, which is an $L\times C$ matrix, fulfils the above conditions.  
 In particular, the continuity condition (i.e., the existence of unique best replies) is ensured for any value of the interference matrix by the  list of preferences set out to reduce power consumption and  inter-team interference.     
\endIEEEproof
As a further remark to the above result, it is worth stressing that the cost introduced in Eq.~(\ref{eq:fullcost}) is an important function that determines whether the game is of strategic complements or substitutes. Indeed, if we consider the payoff to coincide with the utility function (i.e., $\xi=\delta= 0$),  a team's best reply will be to increase its transmit power as the interference grows, implying that the game is of strategic complements. This would lead to an NE in which all teams transmit at maximum power level, without consideration for the interference caused. 
Instead, imposing some $\xi>0$, the game will  turn into a game of strategic substitutes. This is because the first term of the cost function is linear with the received power, and hence increasing with the chosen strategies. Therefore, the payoff function will start decreasing once the increase in the chosen transmit powers does not justify the price the team has to pay. 
Imposing some $\delta>0$ (i.e., activating the second cost component), the relationship between transmit power and cost becomes more complicated but it does not change the nature of the game. The fraction of unserved UEs within the team will be high 
for 
very low power strategies, then it will decrease as the transmit power is increased, and increase again as the strategies chosen cause high intra-team interference. In other words, the second cost component  strengthens the trend in the payoff function imposed by the utility for increasing interference in presence of  low power strategies. For those mid-level strategies that ensure good coverage, it does not affect the cost function.  Instead, it resembles the behaviour of the first cost component for high power strategies, as it is still able to discriminate against high power strategies that may harm the system performance.  

Main results from~\cite{pa-potential,pa-strategic} and references therein show that games of strategic complements/substitutes with aggregation belong to the class of potential games, specifically to the subclass  of {\em{pseudo-potential games}}. 
These games admit pure Nash Equilibria (NE), i.e., action profiles that are a consistent and stable prediction of the outcome of the game, in the sense that no player has incentive to unilaterally deviate from such strategies. 
Another important result that holds for such games with a discrete set of strategies is that, thanks to the continuity condition in Eq.~(\ref{eq:cond-2}),  convergence to an NE is ensured by best reply dynamics \cite{pa-strategic,pa-potential}. 

\section{The power setting algorithm\label{sec:algo}}

We now use the above model and results to build a distributed, low-complexity scheme 
that enables efficient downlink power setting on each CC.
We first consider 
a single carrier  and show how 
the system converges to the best game solution among the possible ones. We then extend the algorithm to the multiple-carrier case and discuss its complexity.

\subsection{Single-carrier scenario\label{subsec:single-carrier}}
Let us first focus on a  single carrier and consider two possible borderline strategies that a team may adopt: the {\em max-power} strategy in which all locations transmit at the highest power level, and the {\em min-power} strategy in which all locations transmit at the lowest available power level greater than 0. Evaluating the utility values obtained for the two extreme strategies, both at the global and individual team level, it transpires that the {\em min-power} always outperforms the {\em max-power} in a HetNet scenario. Indeed, the inter-tier and inter-team interference seriously undermine the overall network performance in terms of global utility, expressed as the sum of all individual team utilities (see  Eqs.~(\ref{eq:SINR})--(\ref{eq:team-utility-sigmoid}) as well as the results in Fig.~\ref{fig:ca-mamihist} in Sec.~\ref{sec:peva}). With regard to the cost, as discussed in Sec.~\ref{subsec:game-analysis}, the first component increases with the increase in the selected transmit power. The second component strengthens the trend imposed by the first cost component  for the {\em max-power} strategy, and by  the utility for the {\em min-power} strategy. This leads to the following important result. 

\begin{theorem}
{\em When  multiple NEs exist, then the NE with the least overall power cost will be the preferred NE in terms of global 
 payoff. This NE will always be reached if players start by setting their strategies to the lowest power level available.}
\end{theorem}
\IEEEproof
The sigmoid function is characterised by a jump reaching a saturation point. 
Since the power cost linearly increases with power, a team's best reply will coincide with the lowest strategy that reaches saturation. Let us assume then that the game has two NEs, in one of which teams tend to choose higher power levels. Since all teams are playing their best replies, they are at utility saturation. Thus, playing higher power level does not ensure higher utility, however it increases the cost component, hence the payoff will be lower in the NE with the higher overall transmitted power. 

A longer, more formal proof can be found in~\cite{techpaper}.
\endIEEEproof

We therefore devise the following procedure that should be executed by each team leader (macro BS), in order to update the BSs downlink power setting, either periodically or upon changes in the user traffic or propagation conditions.  
At a given update period, all teams initialise their transmit power to zero. Then, they  sequentially run the 
Best-reply Power Setting (BPS) algorithm reported in  Alg.~\ref{alg:single-cc-br}. We refer to the single execution of the BPS algorithm by any of the teams as an iteration. Note that the order in which teams play does not affect the convergence or the outcome of the game, since all teams  start from the zero-power strategy. At each iteration, the leader of the team that is playing 
determines the strategy (i.e., the power level to be used at each BS in the team) that represents the best reply to the strategies selected so far by the other teams. The team leader will then notify it to the neighbouring team leaders that can be affected by this choice. BPS  will be run by the teams till convergence is reached, 
which, as shown in Sec.~\ref{sec:peva}, occurs very swiftly. Also, we remark that the strategies identified over the different iterations are not actually implemented by the BSs.  Only the  strategies representing the game outcome will be implemented by the BSs, which will set their downlink power accordingly for the current time period. 

In order to detail how the BPS algorithm (Alg.~\ref{alg:single-cc-br}) works, let us consider the generic $i+1$-th iteration and denote the team  that is currently playing by $t$. The algorithm requires as input the carrier $c$ at disposal of the BSs and the strategies selected so far by the other teams, $\boldsymbol{s^{-t}_c}(i)$. Additionally, it requires the cost components weights $\xi$ and $\delta$, the SINR threshold  $\gamma_{min}$, used to qualify unserved users, and  the utility function parameters $\alpha$ and $\beta$. 
This latter set of parameters are calculated offline and provided to the teams by the network operator. 
The algorithm  loops over all  possible strategies in the strategy set of team $t$, $\boldsymbol{S^t_c}$. For each possible strategy, $\boldsymbol{s}$, and each location $l$ within the team, it evaluates the interference experienced by the tiles within the location area  (line~\ref{line:scc-interference}). This value is used to calculate the SINR and the utility (lines~\ref{line:scc-sinr}-\ref{line:scc-util}), then the first cost component is updated (line~\ref{line:power-cost}).  
In line~\ref{line:quality-cost1}, it is verified whether UEs in tile $z$ achieve the minimum SINR value. 
If not, the cost component $e_t$ is amended to include the affected UEs. 
The overall team utility for each potential strategy 
$\boldsymbol{s}$ is obtained by summing over the individual tile utilities weighted  by the fraction of UEs present in each tile. We recall that such weight factor ensures that the UE distribution affects the outcome of the game accordingly. Once the utility and cost  are obtained, the team payoff corresponding to strategy $\boldsymbol{s}$ is calculated (line~\ref{line:payoff}). After this is done for all possible $\boldsymbol{s}$, the leader  chooses the strategy $\boldsymbol{s^t}(i+1)$ that maximises the team payoff. Note that, according to our game model, the  $\arg\max^{\star}$ function in line~\ref{line:max} operates as follows: in case  the $\arg\max$ function  returns more than one strategy, the leader  applies the list of preferences reported in  Sec.~\ref{subsec:game-definition} to choose the best strategy. 


\begin{algorithm}
\begin{algorithmic}[1]
\Require $c$, $\boldsymbol{s^{-t}_c}(i)$, $\xi,\delta,\alpha,\beta,\gamma_{min}$ \label{line:scc-input}
\ForAll{$\boldsymbol{s}\in\boldsymbol{S^t_c}$}\label{line:scc-str}
\State Set $u^t(\boldsymbol{s},\boldsymbol{s^{-t}_c}(i))$,$w^t(\boldsymbol{s},\boldsymbol{s^{-t}_c}(i))$,$\pi^t(\boldsymbol{s},\boldsymbol{s^{-t}_c}(i))$,$e_t$ 
\hspace{-1mm} to \hspace{-1mm} 0 
\ForAll{$l\in\Lc_t$ {\bf and} $z\in \Zc_l$}
\State Compute $I^t_{z,c}$ by using Eq.~(\ref{eq:interference}) \label{line:scc-interference}
\State Compute $\gamma^t_{z,c}$ by using Eq.~(\ref{eq:SINR}) \label{line:scc-sinr}
\State $u^t(\boldsymbol{s},\boldsymbol{s^{-t}_c}(i)) \hspace{-1mm}\gets u^t(\boldsymbol{s},\boldsymbol{s^{-t}_c}(i))+\hspace{-1mm}\frac{E_z}{E_t\left(1+e^{-\alpha(\gamma^t_{z,c}-\beta)}\right)}$ \label{line:scc-util}
\State $\pi^t(\boldsymbol{s},\boldsymbol{s^{-t}_c}(i))\gets \pi^t(\boldsymbol{s},\boldsymbol{s^{-t}_c}(i))+\xi \bar{a}_{l,c}s_{l,c}$ \label{line:power-cost} 
\If{$\gamma^t_{z,c}\leq \gamma_{min} $} \label{line:quality-cost1}
\State $e_t\gets e_t+\frac{E_z}{E_t}$ \label{line:quality-cost2}
\EndIf
\State $\pi^t(\boldsymbol{s},\boldsymbol{s^{-t}_c}(i))\gets \pi^t(\boldsymbol{s},\boldsymbol{s^{-t}_c}(i))+\delta e_t)$ \label{line:power-cost2}
\EndFor
\State $w^t(\boldsymbol{s},\boldsymbol{s^{-t}_c}(i))\gets u^t(\boldsymbol{s},\boldsymbol{s^{-t}_c}(i))-\pi^t(\boldsymbol{s},\boldsymbol{s^{-t}_c}(i))$  \label{line:payoff}
\EndFor \label{line:scc-strend}
\State $\boldsymbol{s^{t}_c}(i+1)\gets \arg\max^{\star}_{\boldsymbol{s}}w^t(\boldsymbol{s},\boldsymbol{s^{-t}_c}(i))$\label{line:max}
\end{algorithmic}
\caption{\label{alg:single-cc-br}BPS Algorithm run by team $t$ at iteration $i+1$}
\end{algorithm}
 

\subsection{Multi-carrier scenario}

We now extend the previous procedure to the multi-carrier case.  As mentioned before, the team leader  has to decide on the power level to be used at each available carrier, at each location within the team. Thus the team strategy is no longer a vector, but an $L\times C$ matrix, each entry $(l,c)$ indicating the power level to be used for carrier $c$ at location $l$. 
A straightforward extension of Alg.~\ref{alg:single-cc-br} would imply that lines \ref{line:scc-str}--\ref{line:scc-strend} are executed for each element in the new extended strategy set. However, the new strategy set, depending on the number of carriers, may become too large and therefore make the algorithm  impractical to use in realistic scenarios. 

Analysing the utility expression obtained in Eq.~(\ref{eq:team-utility-sigmoid}), we can note that since the carriers are in different frequency bands and have separate power budgets (as foreseen in LTE-A), the utilities secured at each carrier are independent of each other. In other words, the utility a team will get at one of the carriers, is not affected by the strategy chosen at another carrier. The same holds for the first cost component in Eq.~(\ref{eq:fullcost}).  
It is only the second cost component  
that  couples the power setting  at the different carriers. 
Indeed, in networks with carrier aggregation support, a UE can be considered unserved only if the SINR it experiences is below the threshold in all  carriers. 

In order to obtain a practical and effective solution in the multi-carrier scenario, we take advantage of the partial independence between the carriers, and run Alg.~\ref{alg:single-cc-br}  independently for each carrier, keeping the size of the strategy set the same as in the single-carrier scenario. Then, to account for the dependence exhibited by the second cost component, we set  the order in which the per-carrier games are played, using the order of preferences listed in the game description. Since the teams prefer to use high-frequency carriers over low-frequency ones, due to their smaller interference impact, it is logical that the game is played starting from the highest-frequency carrier. It follows that  low-frequency carriers will likely be used to ensure coverage to UEs not served otherwise. 

Importantly, our algorithm is still able to converge to an NE, since surely none of the teams  will deviate from the strategies they chose at each carrier. Also, since the game for the lowest frequency carrier is played last, the number of served UEs  cannot be further  improved without increasing the power level on the other carriers, which we already know is not a preferable move as it has not been selected earlier.  Thus, although it does not search throughout the entire solution space as for the single-carrier scenario, the  procedure is still able to converge to an NE that 
provides a close-to-optimum tradeoff among throughput, user coverage and power consumption.  The results obtained in toy scenarios (see Sec.~\ref{sec:peva}) confirm that our scheme provides performance as good as that achieved by an exhaustive search in the strategy space. 

\subsection{Complexity}
The complexity of the algorithm depends largely on the size of the strategy sets that are available to the teams, $\boldsymbol{S^t}$, since each team has to find the strategy which maximises its payoff value by searching throughout the entire set. The set size depends on the number of discrete power levels available to the BSs ($|\boldsymbol{P}|$), the number of locations in the team ($L$) and the number of CCs available at each location ($C$). In the single-carrier scenario, we have $|\boldsymbol{S^t}|=|\boldsymbol{P}|^L$, while in the multi-carrier scenario the size exponentially grows to $|\boldsymbol{S^t}|=|\boldsymbol{P}|^{LC}$, which is reduced to $|\boldsymbol{S^t}|=C|\boldsymbol{P}|^L$ by our approach. 
\section{Performance evaluation\label{sec:peva}}

We consider the realistic two-tier LTE HetNet scenario that is used within 3GPP for evaluating LTE networks~\cite{scenario}. The network is composed of 57 macrocells and 228 microcells.  Macrocells are controlled by 19 three-sector macro BSs,  while  micro BSs are deployed over the coverage area so that there are 4 non-overlapping microcells per macrocell. The inter-site distance is set to 500~m. The overall network area is divided into 2,478 square tiles of equal size. The BSs are grouped into 57 five-player teams, each consisting of 1 macro BS and 4 micro BSs within its macrocell.   There are about 34,400 UEs in the area, distributed  non-uniformly with a user density around  micro BSs that is three times higher than over the macro BS coverage area. All UEs are assumed to be CA enabled. BSs can use three CCs, each $10$ MHz wide, with the central frequencies: 2.6 GHz (CC1), 1.8 GHz (CC2) and 800 MHz (CC3). The signal attenuation and losses follow the ITU specification for urban environments~\cite{itu}, while the SINR values are mapped to throughput using the look-up table in \cite{sinr-map}. The maximum transmit powers for macro  and micro BSs are set at $20$~W and $1$~W, respectively. The set of discrete power levels is given by $\boldsymbol{P}=\{0,0.1,0.2,...,1\}$, each  representing a fraction of the maximum power. The game is played by all teams using the algorithm for the multiple-carrier scenario. 
The sigmoid function parameters are $\alpha=1$ and $\beta=1$, which were selected as the most appropriate to model the relationship between the selected strategy and final user rate. 
The SINR threshold is set at $\gamma_{min}=-10$~dB, based on 
 \cite{sinr-map}. 

Using our results in \cite{techpaper}, the value of the cost parameter is set as $\xi= \frac{k\alpha}{\bar{\mathbf{I}}}$,  where $k$ is the weight factor used to indicate the importance we place on the first cost component and $\bar{\mathbf{I}}$ is an average  value for interference calculated by the network operator, obtained by fixing the transmit power of all teams at half the maximum power. 
Unless otherwise specified, the weight factor $k$ is set  to $0.25$ while 
$\delta=0.6$. These values were selected based on their effect on the performance metrics, as shown below in the simulation results. 

The performance of the algorithm is first compared to the optimum in a toy scenario. In the large-scale scenario described above, it is instead compared to that of four baseline strategies: 
the two fixed power strategies {\em max-power} and {\em min-power}, as well as to the {\em max-power} strategy  coupled with eICIC technique, as  usually applied in the literature and in  practice, and with the Low Power - ABS (LP-ABS) technique \cite{lp-abs}. Traditional eICIC is applied with CRE for microcells set at $8$~dB and 
macro BS downlink transmissions muted in  25\% of subframes (ABS). These values were chosen to represent the mid-range of those applied in the surveyed literature \cite{eicic-alg} 
LP-ABS uses a $6$~dB microcell biasing, ABS subframe ratio of $50\%$ and macro BS power reduction of $6$~dB during ABS, which were shown to perform best in \cite{lp-abs}. Note that for the strategy reached via the BPS algorithm and the two fixed power strategies, user association is distance-based and fixed, while for the power strategies coupled with eICIC, it is based on the strongest received pilot signal plus the bias, to properly model the CRE behaviour.

\begin{figure}
\centering
\includegraphics[width=0.23\textwidth]{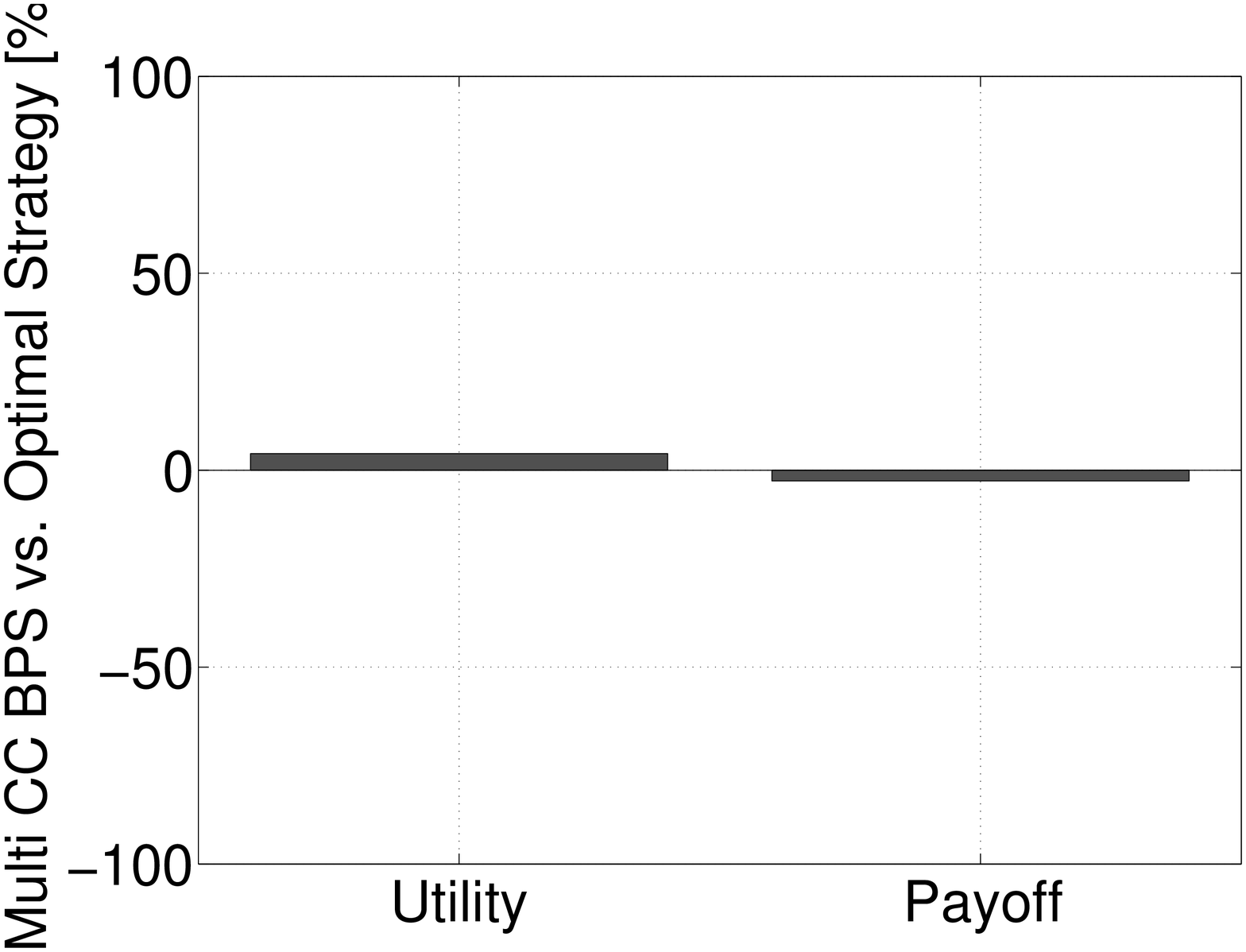}
\includegraphics[width=0.23\textwidth]{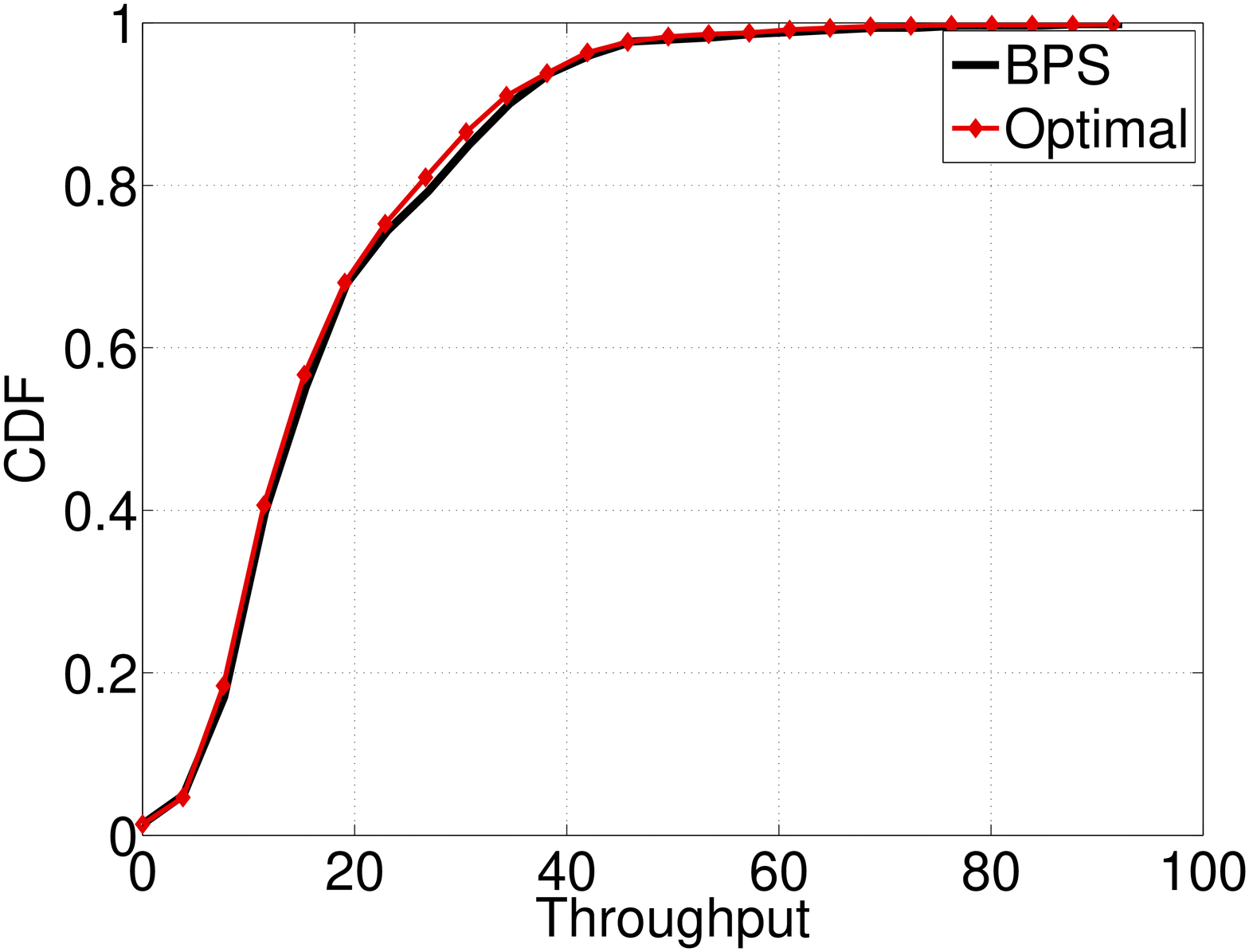}
\caption{\label{fig:ca-comparison}Deviation from optimal strategy: utility, payoff and overall transmitted power (left) and CDF of the per-user throughput (right).}
\vspace{-5mm}
\end{figure}

In Fig.~\ref{fig:ca-comparison} we compare BPS in a multi-carrier setting with the optimal solution obtained via exhaustive search. Due to the problem complexity, the comparison is performed only for a toy scenario in which two teams compete, each consisting of one macro and one micro BS. The results, obtained by averaging the behaviour of ten different sets of teams, show that there is negative deviation in terms of payoff as expected, but BPS yields higher utility. 
Looking at the per-user throughput CDF curves, however, we note that the two strategies perform almost identically.

\begin{figure}
\centering
\includegraphics[width=0.18\textwidth]{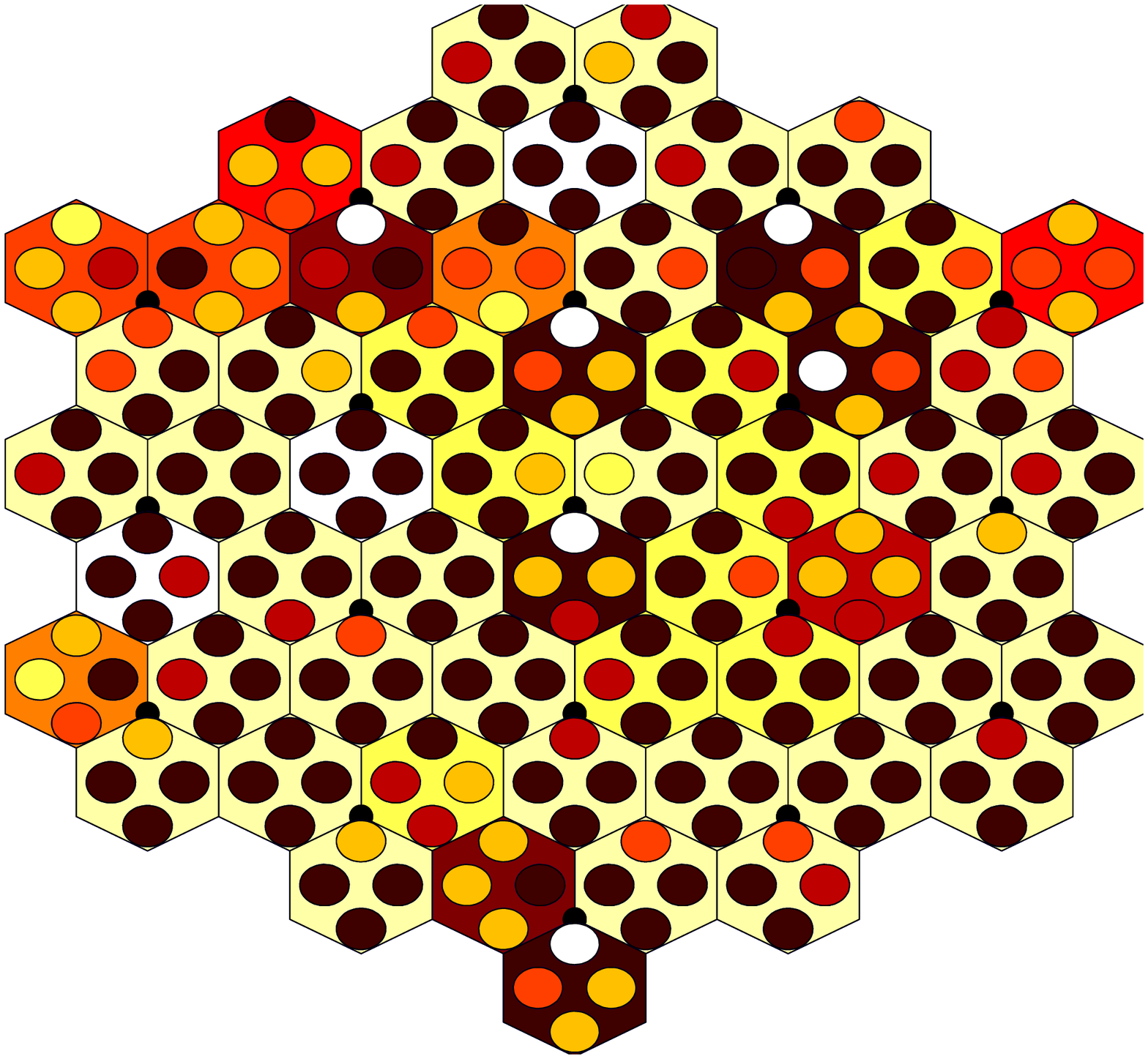}
\hspace*{0.0cm}
\includegraphics[width=0.18\textwidth]{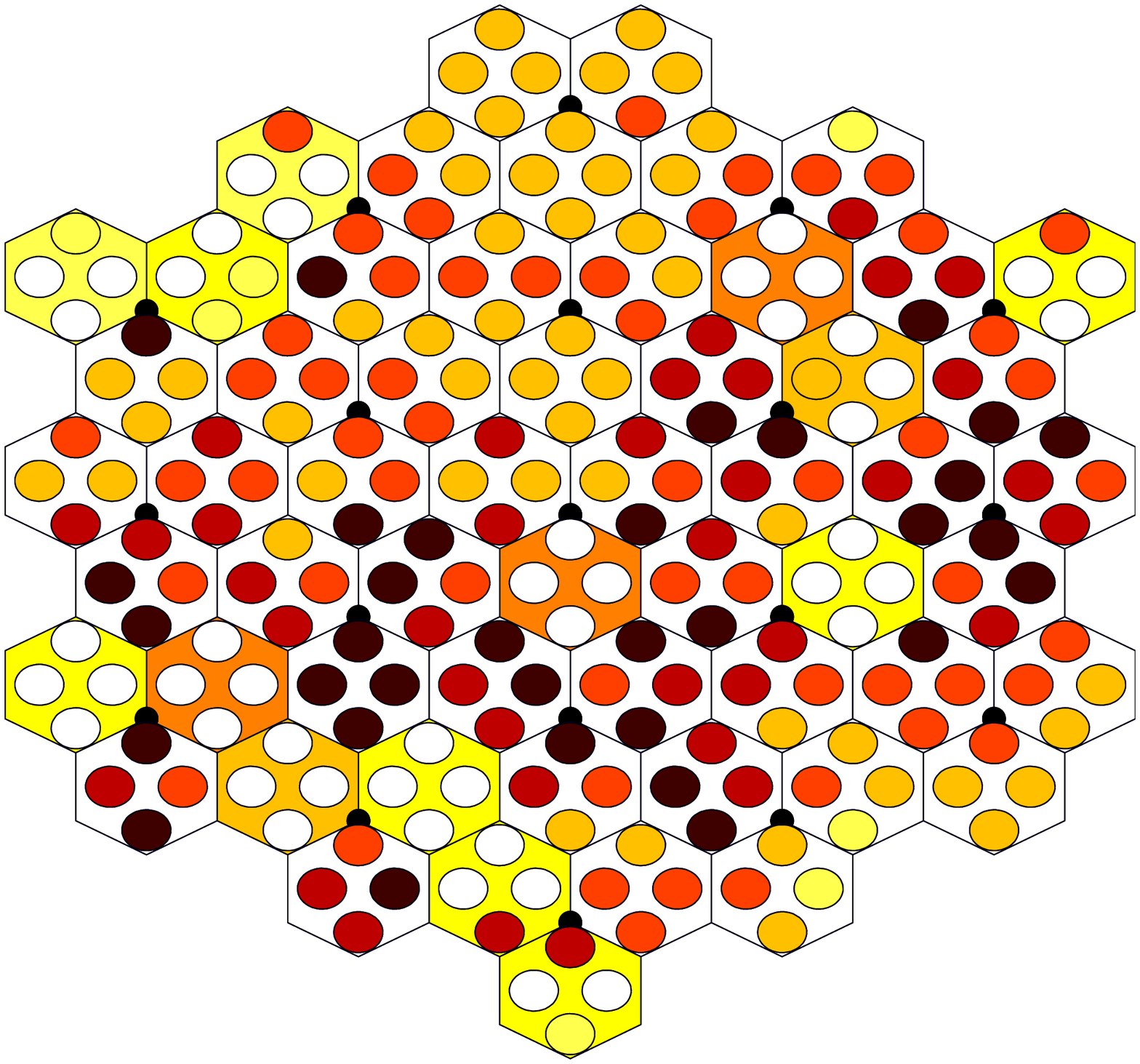}
\includegraphics[width=0.18\textwidth]{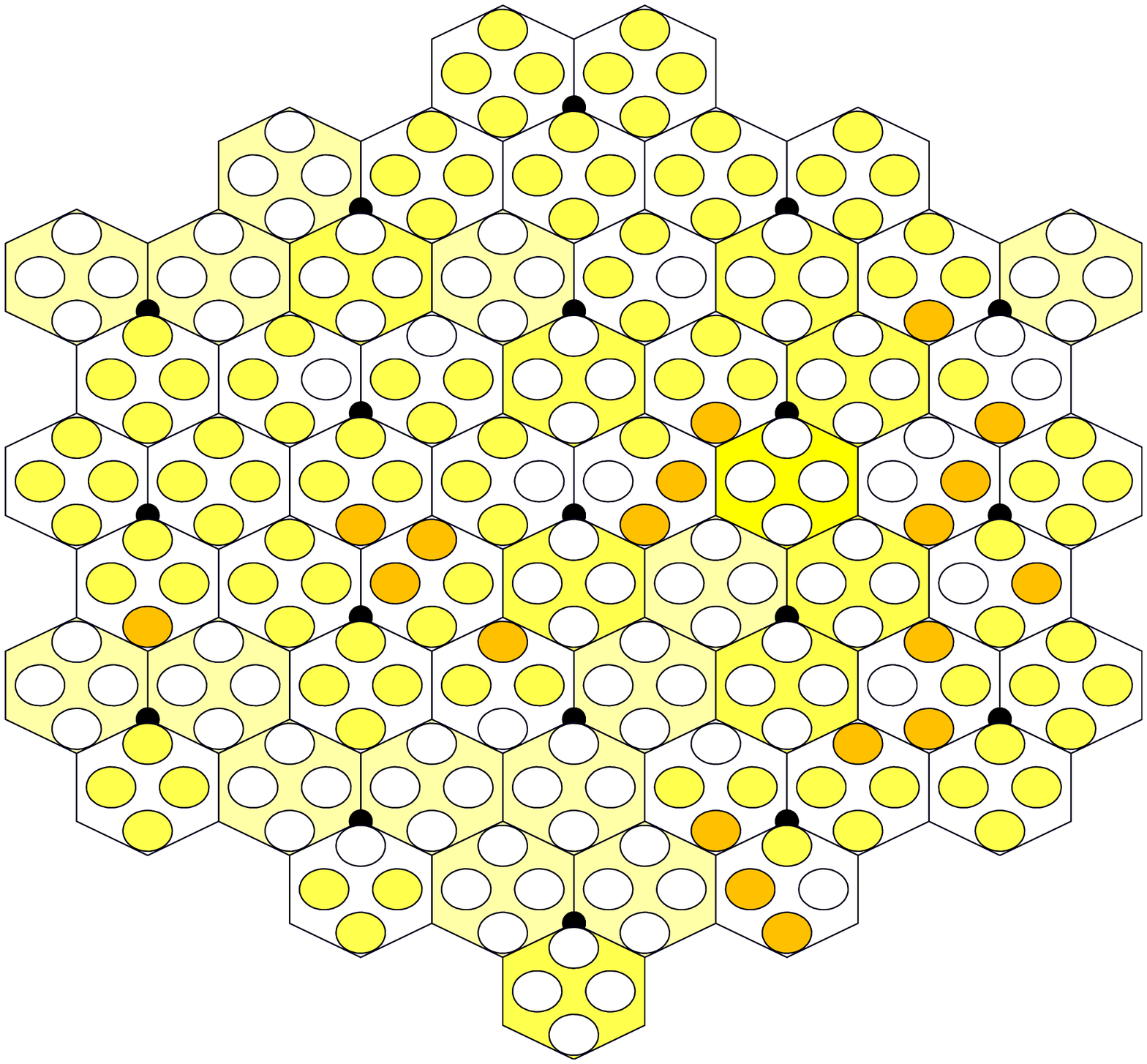}
\caption{\label{fig:ca-cc1-ne}BPS strategies for a 57-team game for CC1 (top left), CC2 (top right) and CC3 (bottom). Darker shades represent higher power level, while the white color corresponds to the {\em off} state. Hexagons  are macro BSs while circles are micro BSs.}
\vspace{-5mm}
\end{figure}

\begin{figure}
\centering
\includegraphics[width=0.23\textwidth]{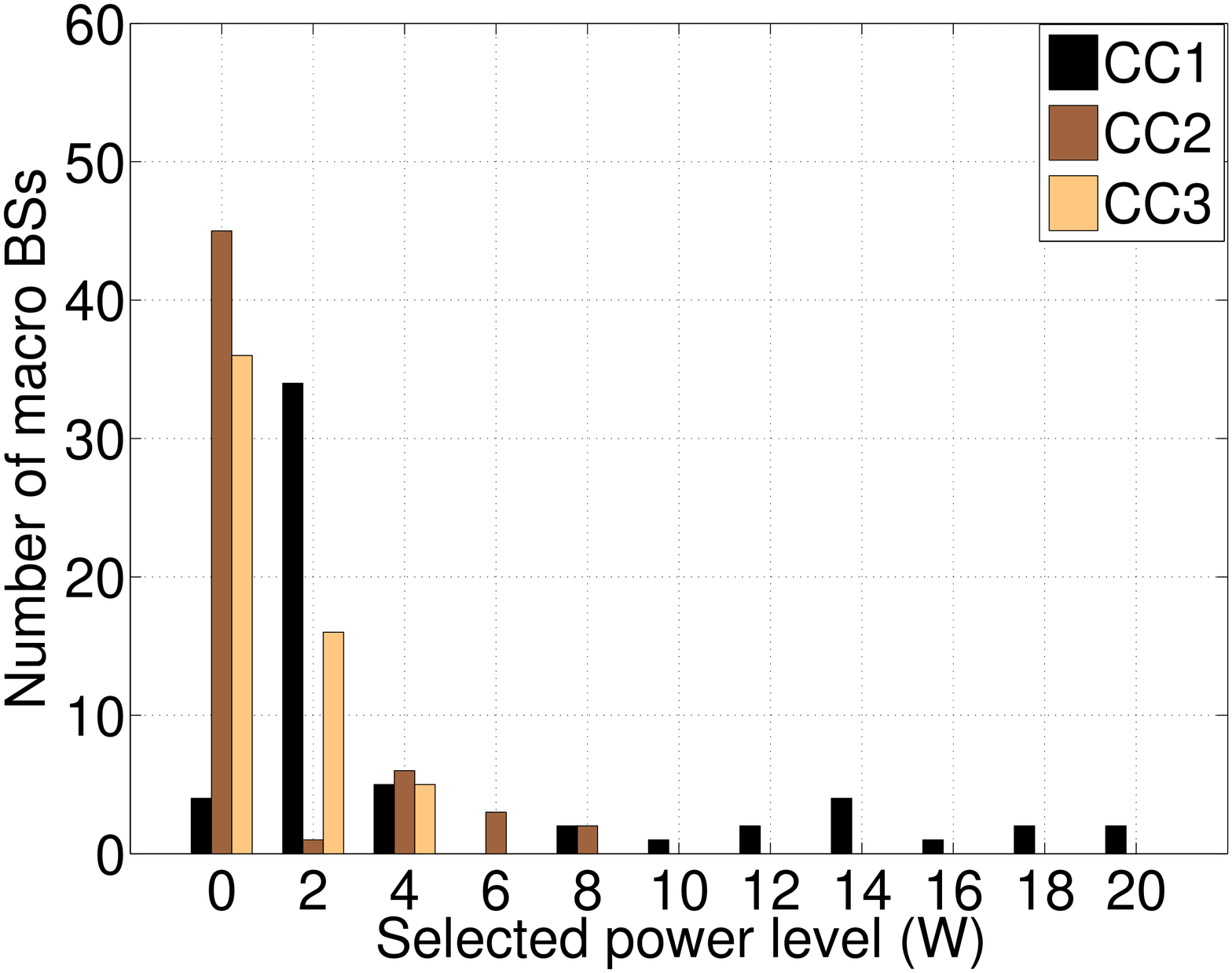}
\includegraphics[width=0.23\textwidth]{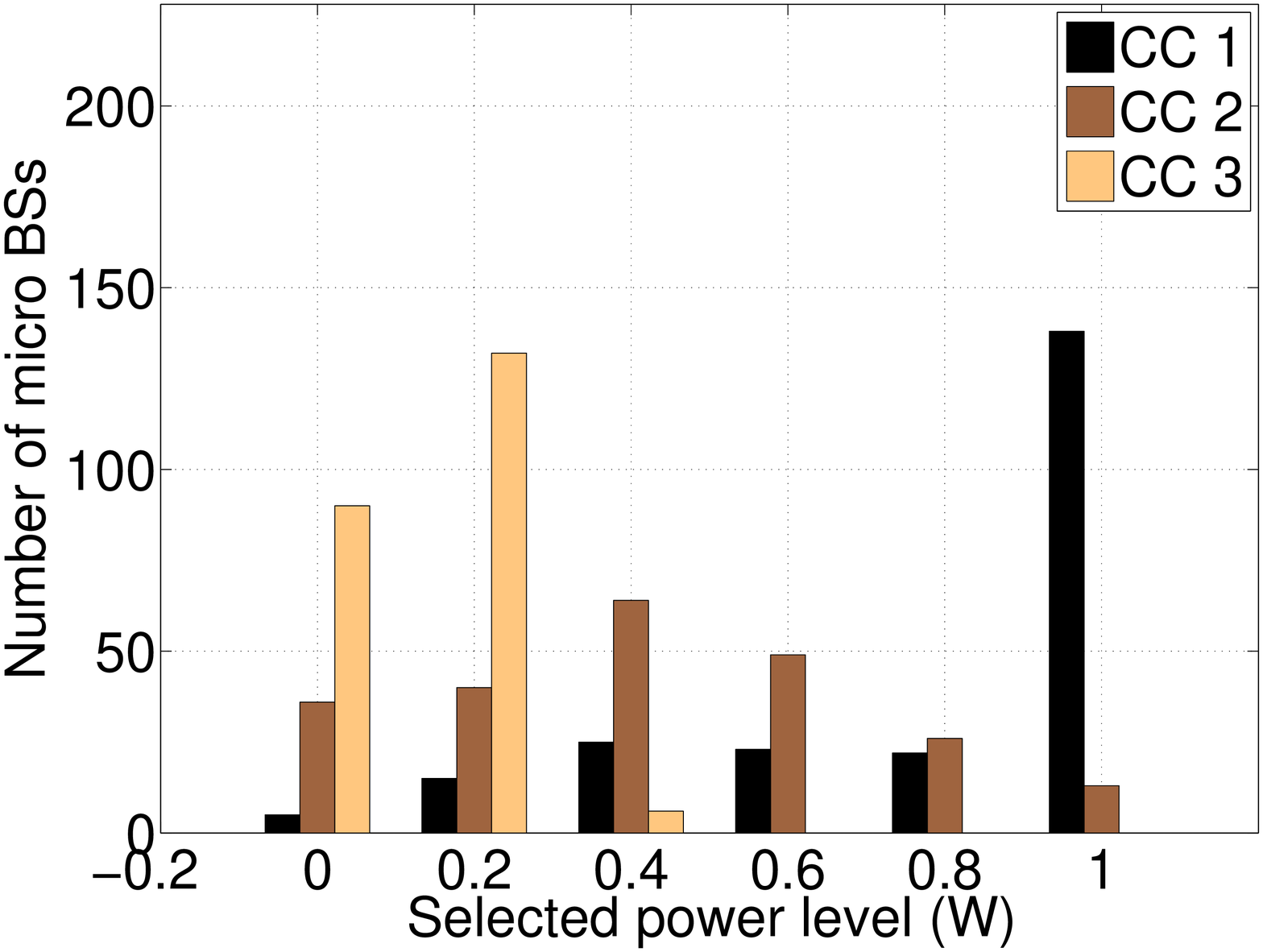}
\caption{\label{fig:ca-mamihist1}BPS strategies for a 57-team game:  chosen strategies by macro (left) and micro (right) BSs.}
\vspace{-5mm}
\end{figure}

In Fig.~\ref{fig:ca-cc1-ne}, we  look at a snapshot of the NE strategy reached via the BPS algorithm, in a game with 57 teams. The strategies chosen by the teams for each CC are differentiated using different shades, from white ({\em zero} power) to black ({\em maximum} power). Hexagons represent the macro BS, while circles represent micro BSs.  The figure shows that CC1, i.e., the high frequency carrier, allows for higher transmit power to be used by both macro and micro BSs, due to its low interference impact. CC1 can be also used simultaneously by macro and micro BSs in the same team, which is not always the case for the other two CCs. CC2 and CC3 are used to complement each other to ensure overall coverage. Histograms of chosen strategies for macro and micro BSs, shown in Fig.~\ref{fig:ca-mamihist1}, confirm these observations. Here note that CC1 is activated for most macro and micro BSs, however macro BSs often set low power levels for CC1, while most micro BSs set CC1 at maximum power level. On the contrary, CC2 is rarely activated for macro BSs, while CC3 (the low frequency CC) is the least utilised, and tends to be especially unfavored by micro BSs, due to its high interfering impact. These results validate the intuition that far reaching low-frequency carriers are not appropriate to be used by micro BSs, rather they should be used only to ensure broader coverage for edge UEs. 

\begin{figure*}[b!]
\centering
\includegraphics[width=0.28\textwidth]{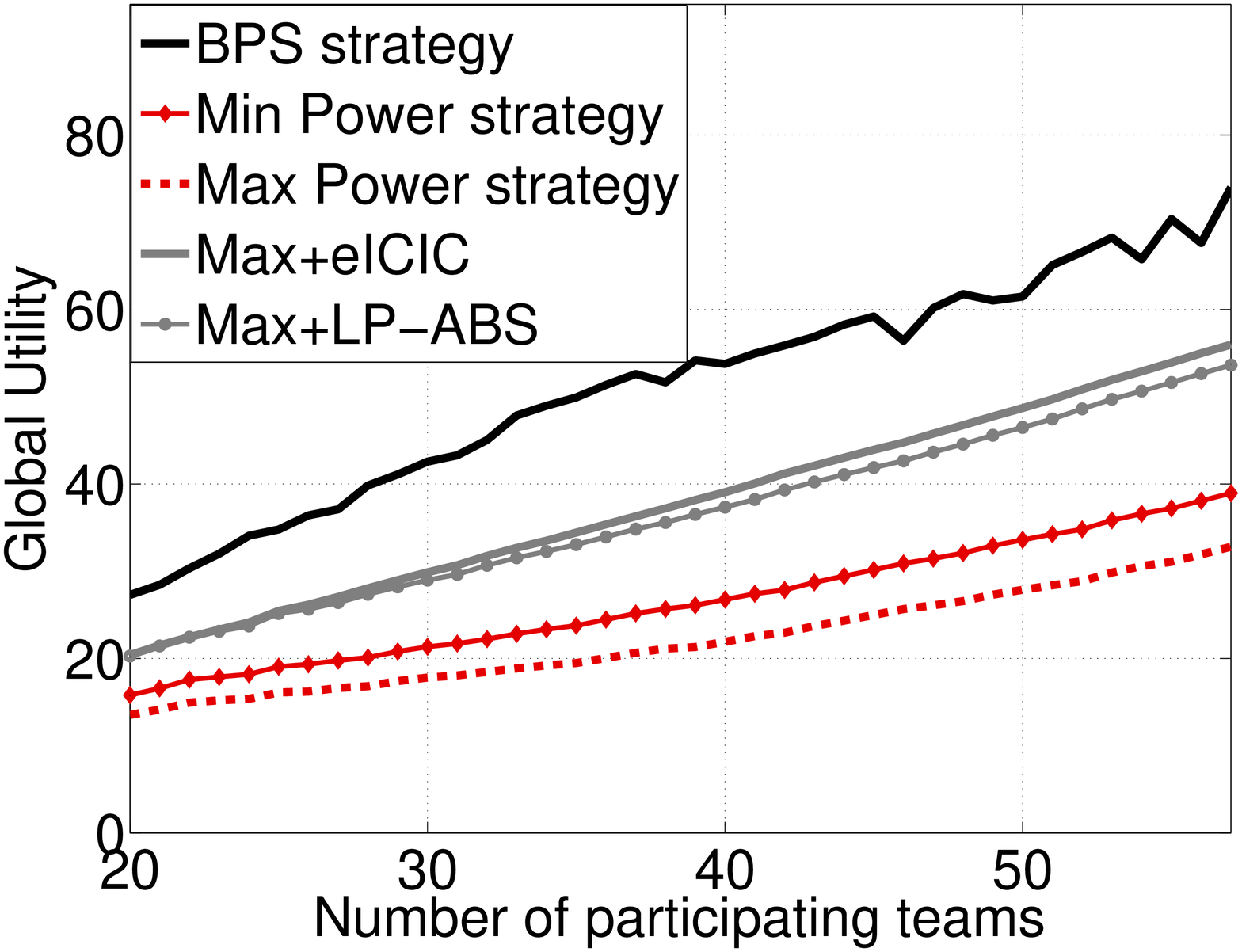}\hspace{2mm}	
\includegraphics[width=0.28\textwidth]{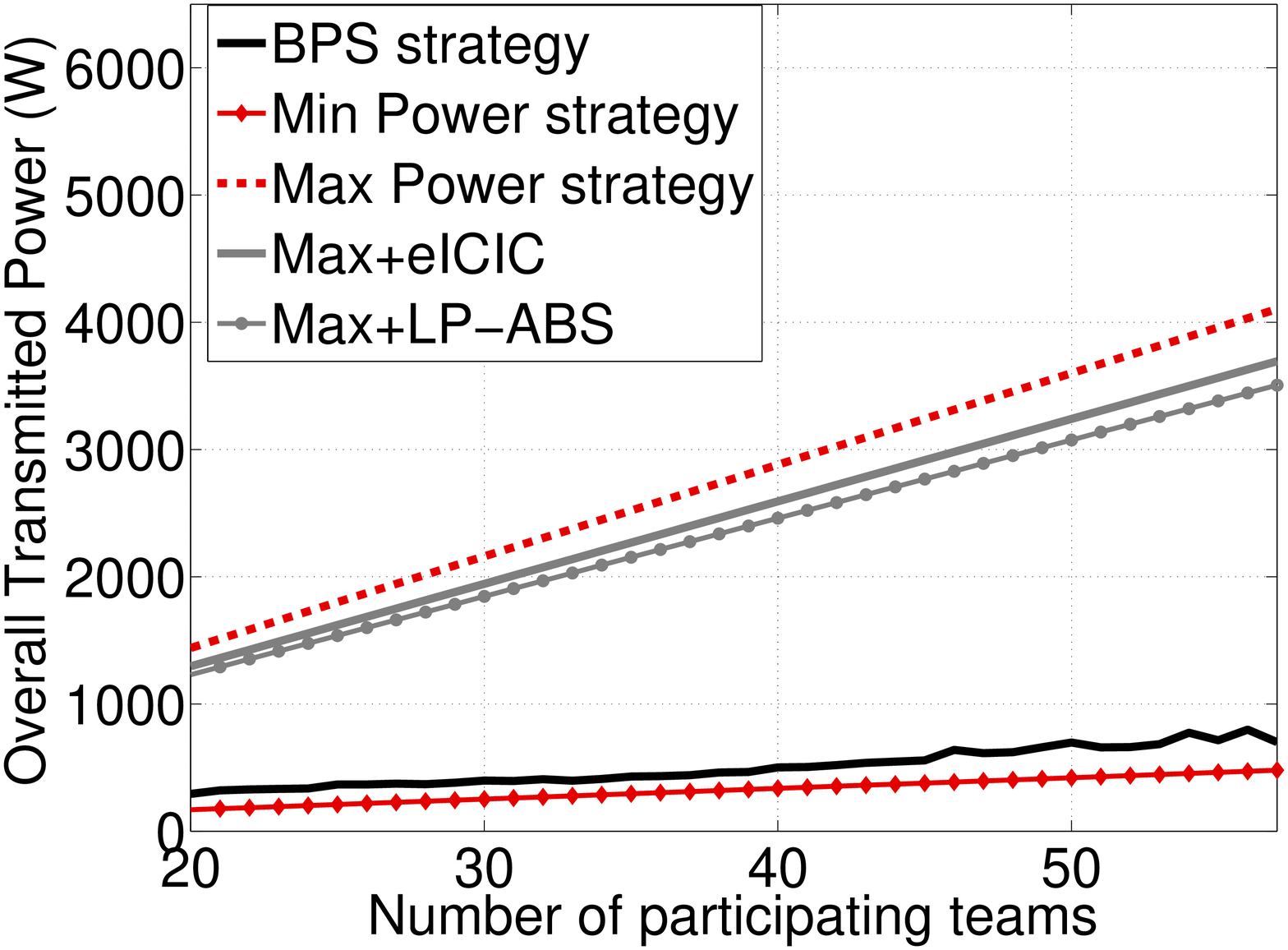}\hspace{2mm}	
\includegraphics[width=0.28\textwidth]{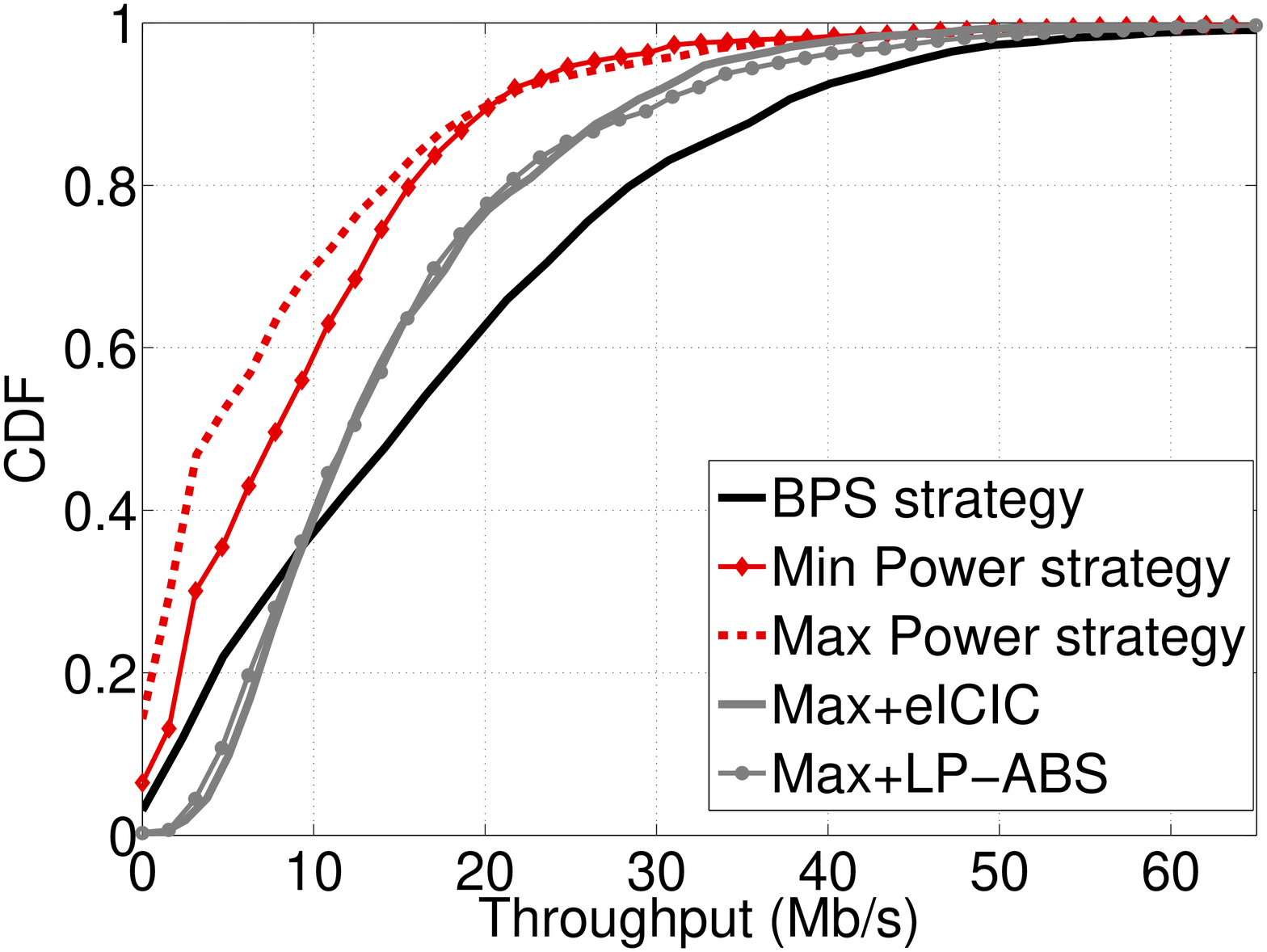}
\caption{\label{fig:ca-mamihist}BPS strategy for a 57-team game: comparison with baseline strategies for varying number of  teams. Global utility (left), overall transmitted power (middle) and CDF of the per-user throughput (right).}
\vspace{-3mm}
\end{figure*}

\begin{figure*}
\centering
\includegraphics[width=0.23\textwidth]{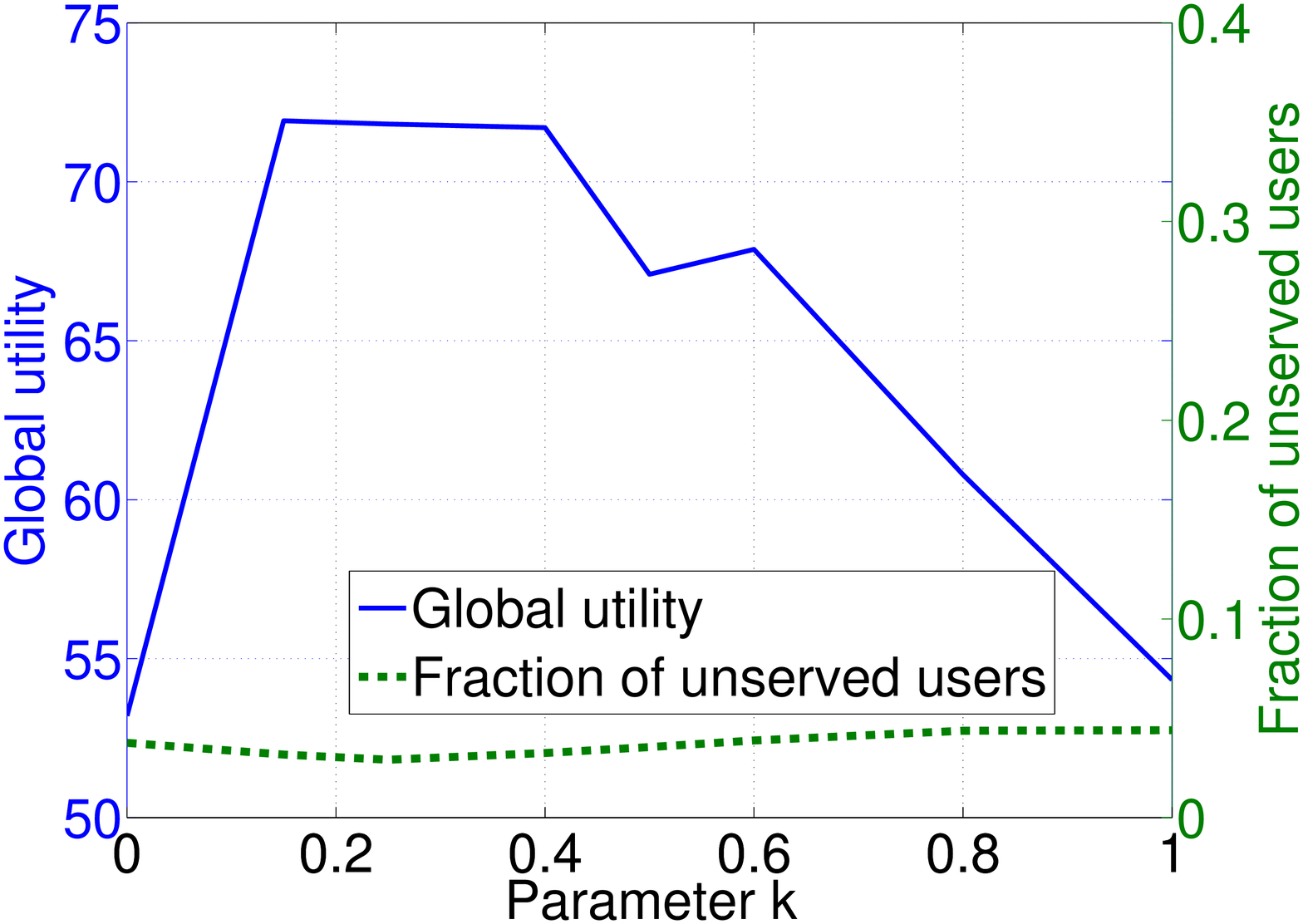}
\includegraphics[width=0.23\textwidth]{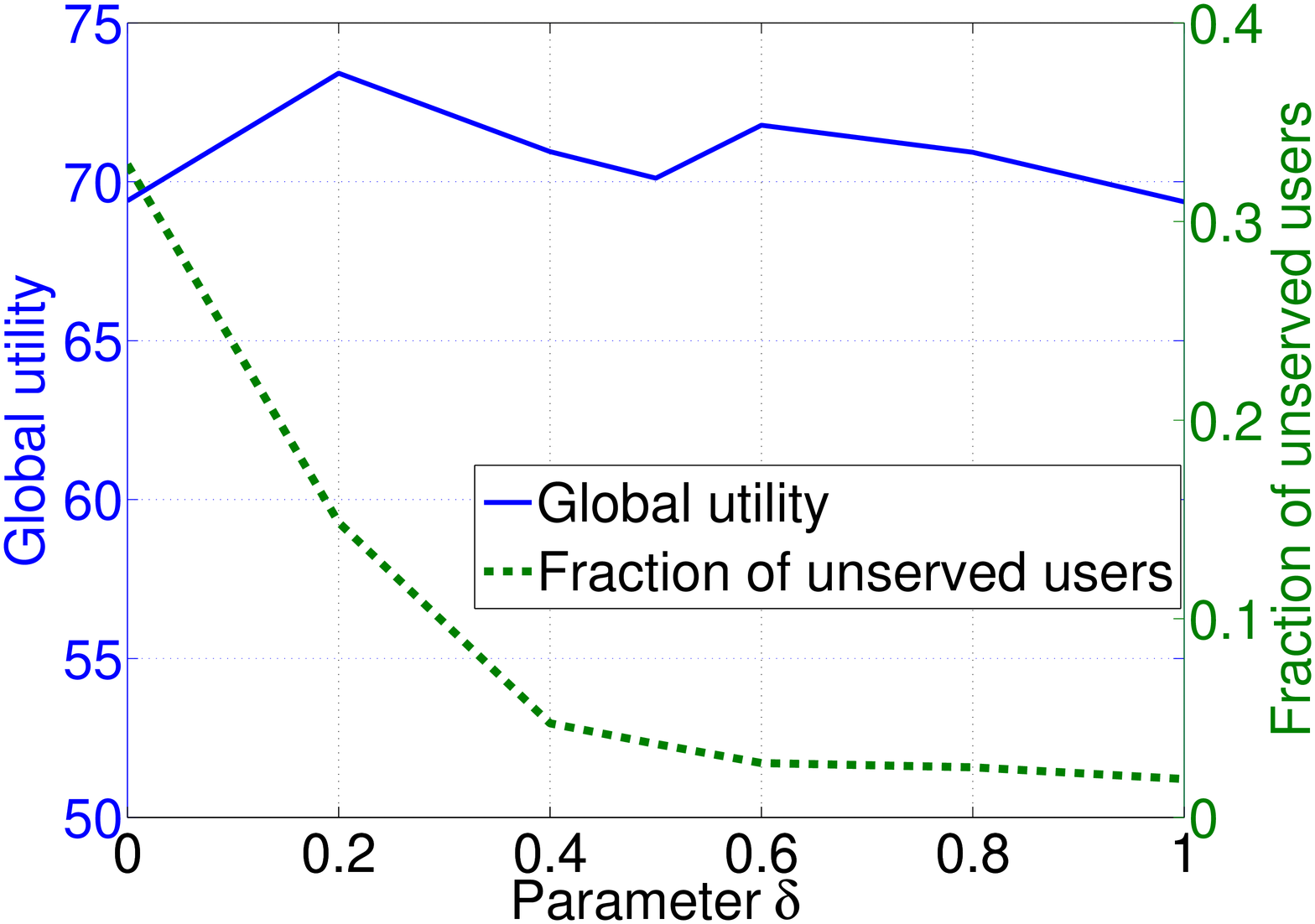}	
\includegraphics[width=0.23\textwidth]{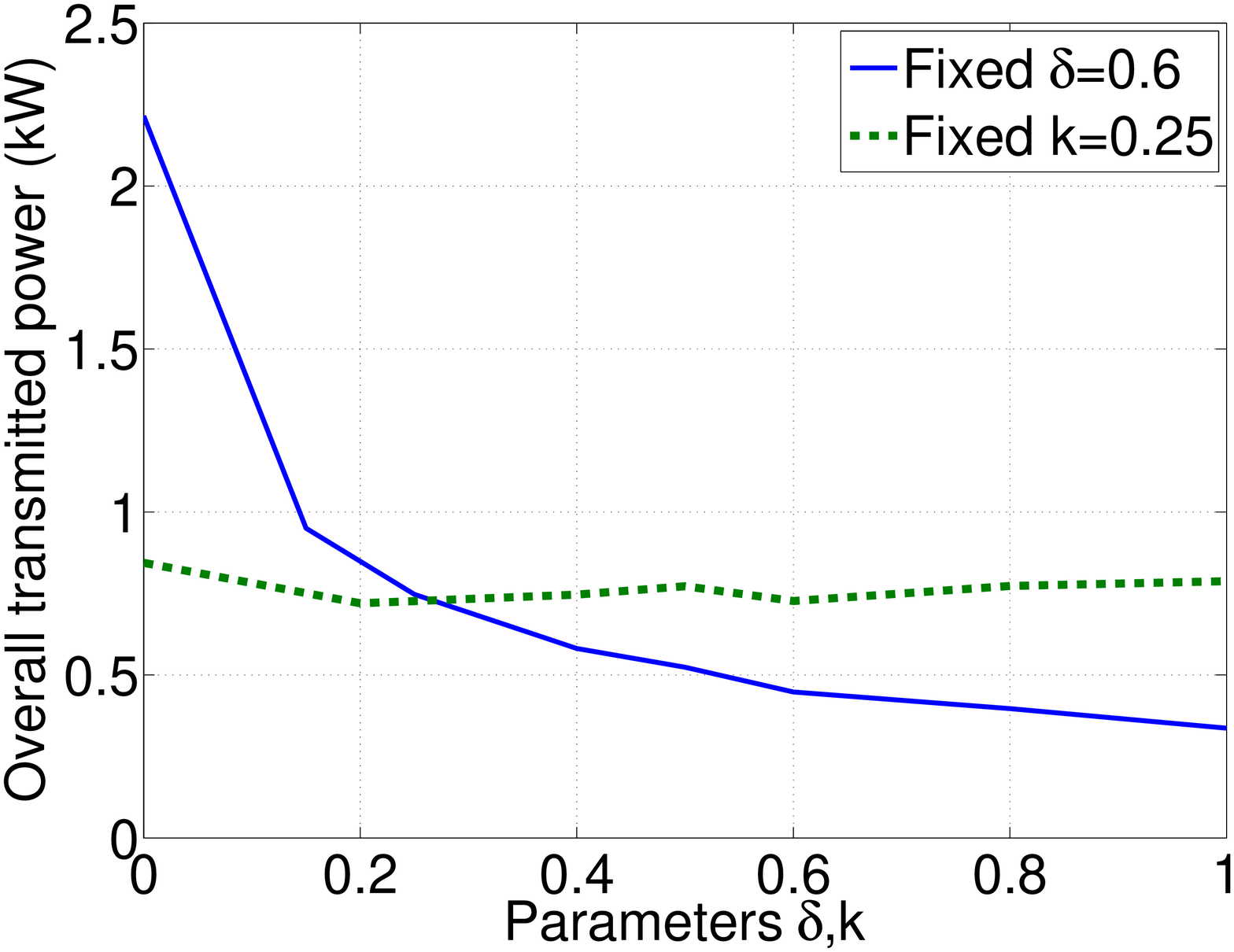}
\includegraphics[width=0.23\textwidth]{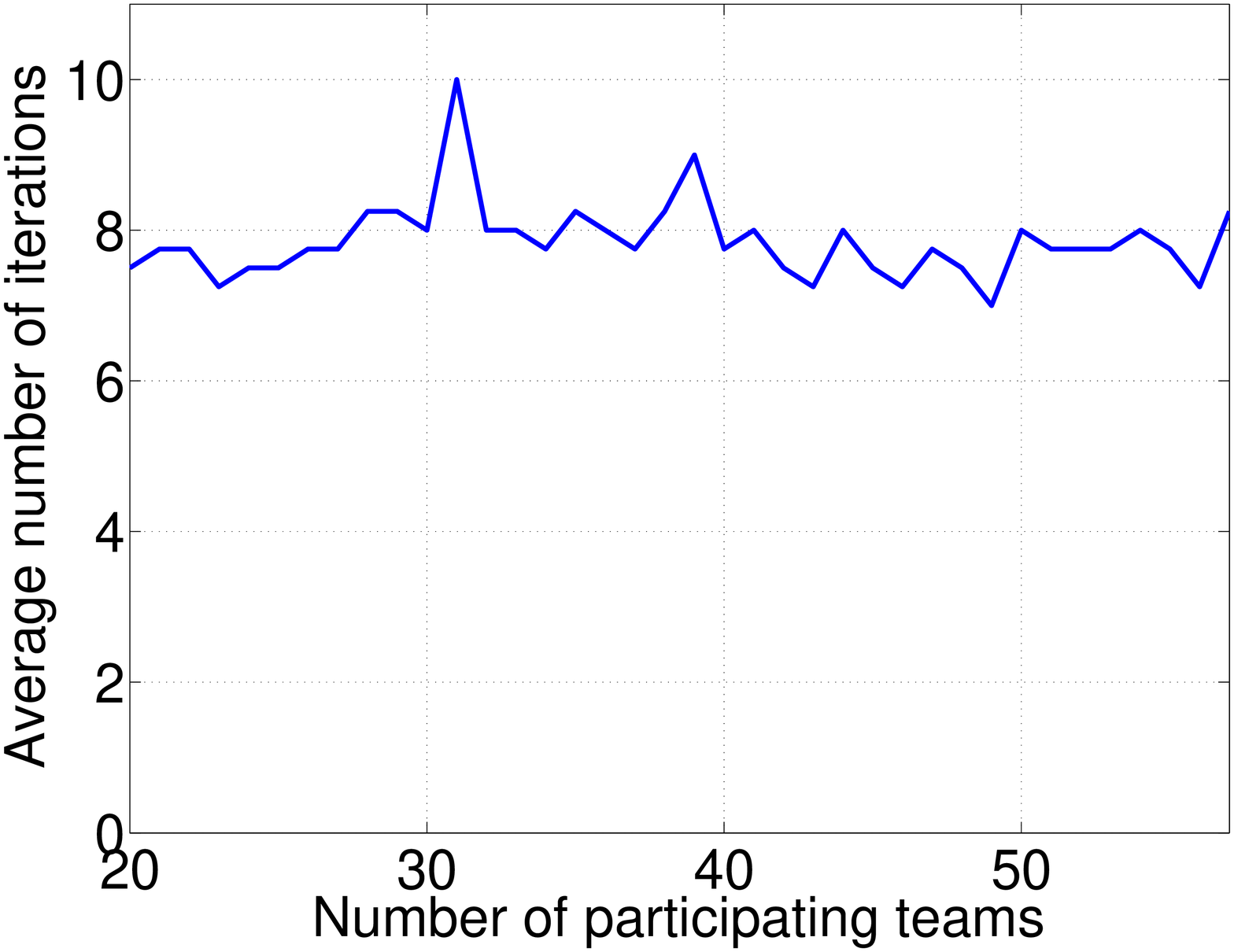}
\caption{\label{fig:ca-priceeval}From left to right:  Effect of the price parameter $k$ on global utility (solid, blue) and fraction of unserved users (dashed, green) with for $\delta=0.6$;  Effect of the coverage cost parameter $\delta$  on global utility (solid, blue) and fraction of unserved users  (dashed, green), with $k=0.25$; Effect of  $k$ (solid, blue) and $\delta$ (dashed, green) on the overall transmitted power; Per-team number of iterations for game convergence vs. number of teams.}
\vspace{-5mm}
\end{figure*}

Next, in the left and middle plots of Fig.~\ref{fig:ca-mamihist} we compare  the performance of the strategy reached via our scheme (labelled by ``BPS'') to the fixed baseline strategies, in terms of global utility and overall transmitted power, and for a varying number of  teams. The strategy reached via the BPS mechanism outperforms all other solutions in terms of global utility, calculated as the sum of the individual team utilities. Also, the gap in performance grows  with the number of  teams. This gain in performance is achieved at much lower transmit power, which implies that the BPS strategy is  very efficient. The overall transmit power of the BPS strategy, calculated as the sum of the selected transmit powers over all BSs and CCs in the network, closely approaches that of  the {\em min-power} strategy and is much lower than the power consumption of all  other schemes. Also, as anticipated in Sec.~\ref{subsec:single-carrier},  the {\em min-power} strategy always outperforms the {\em max-power} strategy in terms of utility, regardless of the number of teams, while keeping the overall transmit power at the minimum level. 

The final comparison is performed in the rightmost plot of Fig.~\ref{fig:ca-mamihist}, which depicts the cumulative distributive function (CDF) of the per-user throughput for the strategies under consideration. 
 Overall, our solution   outperforms all other schemes. This holds especially for the top $70\%$ of UEs. eICIC and LP-ABS give slightly better results in ensuring a positive throughput to the worst  UEs. However, BPS  
provides a very low fraction of UEs that are left unserved (about $2\%$), while  transmitting at much lower overall power. Note also that the strategies with eICIC and LP-ABS are at a slight advantage since user association is performed based on the best downlink pilot signal, which, at least for downlink communication, is always better than the fixed distance-based user association scheme that we assumed for simplicity. In summary, it is clear that BPS is a very well-balanced strategy in terms of level of service: it  provides  slightly lower per-user throughput than  eICIC and LP-ABS for the worst UEs, but much better throughput than all other strategies for the rest of the UEs, and it  consumes very little power (almost the same as the {\em min-power} strategy). 
 
In Fig.~\ref{fig:ca-priceeval}, we look at the behaviour of our algorithm. 
First, we evaluate the effect of $k$, i.e., the weight we assign to the cost of received power, on the global utility and the fraction of low SINR users, by varying its value from $0$ to $1$ and fixing  $\delta=0.6$. We see that increasing $k$ is beneficial in terms of global utility (solid, blue line), but only up to some value (around 0.4). Beyond that, the global utility experiences a sharp drop, which signifies that, due to the high power price, BPS is more inclined to provide strategies that optimise power consumption rather than the utility. Also,  $k$ has little effect on the fraction of unserved users (dashed, green line): just a small improvement can be noticed around $k=0.25$.  Conversely, the cost parameter $\delta$ plays an instrumental role in ensuring that the number of UEs experiencing an SINR below the acceptable threshold is kept low, as can be seen by the dashed green line in  the second plot of Fig.~\ref{fig:ca-priceeval} 
(here $k=0.25$). The third plot depicts  the effect of $k$ (solid, blue line) on the overall transmitted power when $\delta=0.6$, and the effect of $\delta$ (dashed, green line) when $k=0.25$. Note that increasing $k$ leads BPS to converge to strategies with overall lower power, however, as observed before, this comes at the expense of the utility.  As expected, the increase in $\delta$ does not lead to strategies with higher overall transmit power, which confirms our earlier statement that introducing the second cost component does not change the nature of the game. 

Finally, the rightmost plot presents the average number of iterations it takes to each team to converge to the final best strategy. Depending on the intra-team dynamics, teams may take a different time, however the game always  converges quite fast (in about 8 iterations). Importantly, the average number of iterations required by each team does not grow with the number of  teams. 


\section{Conclusions\label{sec:concl}} 
We proposed a novel solution for downlink power setting in HetNets with carrier aggregation, which aims to reduce interference and power consumption, and to provide high quality of service to users. Our approach leverages the different propagation conditions  of the carriers and the different transmit power that macro and micro BSs can use for them. 
Through game theory, we framed the problem as a competitive game among teams of macro and micro BSs, and identified it as a game of  strategic substitutes/complements with aggregation.  
We then introduced a distributed algorithm that enables the teams to reach a desirable NE 
in very few iterations. Simulation results, obtained in a realistic scenario, show that our solution greatly outperforms the existing  strategies in terms of global performance while consuming little power. 

\end{document}